\documentstyle[a4,12pt,epsf,vmargin]{article}
\setmarginsrb{24mm}{28mm}{24mm}{28mm}{0pt}{0mm}{0pt}{0mm}
\begin{document}
\begin{titlepage}
\centerline{\normalsize\bf TOWARDS RECONSTRUCTING THE FINAL STAGE}
\centerline{\normalsize\bf OF HEAVY ION 
COLLISIONS\footnote{Paper presented at the XXXIIth Rencontres de Moriond,
``QCD and High Energy Hadronic Interactions'', Les Arcs, France,
March 22-29, 1997.}}
\vskip 1.0cm
\centerline{\normalsize Urs Achim Wiedemann}
\centerline{\small\it Institut f\"ur Theoretische Physik, 
                      Universit\"at Regensburg,}
\centerline{\small\it D-93040 Regensburg, Germany}
\vskip 7.0cm
\begin{abstract}
{\normalsize 
A Fourier inversion problem lies
at the heart of determining spatio-temporal characteristica of the 
final stage of a heavy ion collision: From the measured two-particle
momentum correlations $C({\bf p}_1,{\bf p}_2)$ of identical
particles, pions say, a Hanbury-Brown /Twiss (HBT) interferometric analysis
aims at extracting as much information as possible about the Wigner phase
space density $S(x,p)$ of pion emitting sources in the collision 
region,~\cite{S73,CH94}
 \begin{equation}
   C({\bf q},{\bf K}) =
   1 + {\left\vert \int d^4x\, S(x,K)\,
   e^{ix{\cdot}q}\right\vert^2 \over
   \int d^4x\, S(x,p_1)\, \, \int d^4y\, S(y,p_2) } \, ,
   \label{1}
 \end{equation}
where $q = p_1 - p_2$, $K = \textstyle{1\over 2}(p_1 + p_2)$. Here,
we discuss how the analysis of (\ref{1}) allows to separate the
effects of temperature and transverse flow which cannot be disentangled
completely on the basis of single-particle spectra.
}
\end{abstract}
\end{titlepage}
%
Physics is rich in Fourier inversion problems
which are not analytically invertible. A well-known example is the 
determination of the charge density distribution $\rho({\bf r})$ of 
atomic nuclei from the structure function $F({\bf q})$ measured in 
elastic electron scattering,
  \begin{equation}
    \label{2}
    F^2({\bf q}) = \vert \textstyle{1\over e} \int d{\bf r}\,
        \rho({\bf r})\, {\rm e}^{i{\bf q}{\bf r}}\vert^2\, .
  \end{equation}
In most cases, this structure function was measured only in a small
window of the momentum transfer ${\bf q}$ which makes an analytical
unfolding of $\rho({\bf r})$ without additional assumptions 
impossible. The standard way out is to take recourse to model 
distributions $\rho({\bf r})$ like e.g. the Fermi distribution 
$\rho({\bf r}) = \rho_0/ 
{\left({ 1 + \exp(\textstyle{{{\bf r} - R}\over z}) }\right)}$ whose
parameters have a simple physical interpretation. In a first step, 
these model parameters are determined in a fit of $F({\bf q})$,
till a refined analysis reveals the need for more detailed models
which include e.g. dips in the charge distribution for 
small ${\bf r}$, cf. ~\cite{H57}.

In our analysis of the measured two-particle correlations (\ref{1}), 
we adopt essentially the same strategy, for different reasons though:
the detected pions which determine $C({\bf q},{\bf K})$ are on mass-shell 
and this results in a constraint for the relative pair momentum $q$, 
$q\cdot K = 0$. Hence, the four-vector $q$ has only three independent 
components and even for an exactly known correlator, the Fourier 
transform (\ref{1}) is not uniquely invertible. This necessitates
a model-dependent approach.

The analysis presented here is based on a model pion emission function
including resonance decay channels $R$: 
  \begin{equation}
     S_{\pi}(x,p) = S_{\pi}^{\rm dir}(x,p) + \sum_{R} S_{R\to \pi}(x,p)\, .
  \label{3}
  \end{equation}
The contributions $S_{R\to\pi}$ are obtained from
the direct resonance emission function $S_R^{\rm dir}(X,P)$  
by propagating the resonances of widths
$\Gamma$, produced at $(X_\mu,P_\mu)$, along a classical path 
$x^\mu = X^\mu + {P^\mu\over M} \tau$ according to an exponential
decay law~\cite{S92,HEI96,WH96}:
  \begin{equation}
   S_{R\to\pi}(x,p) = 
        \int_{\bf R}\, \int d^4X \, 
        \int d\tau \, \Gamma e^{-\Gamma\tau} \,
        \delta^{(4)}\textstyle{\left( x - 
            \left( X + {P\over M} \tau \right) \right)} 
        S_R^{\rm dir}(X,P)\, ,
 \label{4}
 \end{equation}
where $\int_{\bf R}$ is the integral over the available resonance
phase space for isotropic decays.
Our model assumes local thermalization at freeze-out and produces 
hadronic resonances by thermal excitation. For particle species $i$ 
with spin degeneracy $2J_i+1$, the emission function reads \cite{WH96}
 \begin{equation}
 \label{5}
   S^{\rm dir}_i(x,P) = \textstyle{2J_i + 1 \over (2\pi)^3}\,
   P{\cdot}n(x)\,
   \exp{\left(- {P \cdot u(x) - \mu_i \over T} \right)}\,  
          \exp\left( - {r^2\over 2 R^2} 
                     - {\eta^2\over 2 (\Delta\eta)^2}
                     - {(\tau-\tau_0)^2 \over 2 (\Delta\tau)^2}
                 \right) \, . 
 \end{equation}
The Boltzmann factor $\exp[-(P{\cdot}u(x) - \mu_i)/T]$ 
implements both the assumption of thermalization, with temperature $T$ 
and chemical potential $\mu_i$, and collective expansion with 
hydrodynamic flow $4$-velocity $u_{\mu}(x)$. The geometrical extension
of the collision region is determined by Gaussian widths $R$ and 
$\Delta\eta$ in the transverse radius $r$ and the space-time 
rapidity $\eta= {1\over 2} \ln{[(t+z)/(t-z)]}$, as well as
a Gaussian average around a mean freeze-out proper time $\tau_0$ 
with dispersion $\Delta\tau$.
Freeze-out occurs along proper time hyperbolas $P{\cdot}n(x)= M_\perp 
\cosh(Y-\eta)$ where $Y$ and $M_\perp$ are the rapidity 
and transverse mass associated with $P$.
In the longitudinal direction we assume scaling expansion, $v_l=z/t$
or $\eta_l = {1 \over 2} \ln[(1 + v_l)/(1-v_l)] = \eta$, for the
transverse expansion a linear rapidity profile:
 \begin{equation}   
 \label{6}
   \eta_t(r) = \eta_f \left({r\over R}\right)\, .
 \end{equation}
We include all pion decay channels of $\rho$, $\Delta$, $K^*$, 
$\Sigma^*$, $\omega$, $\eta$, $\eta'$, $K_S^0$, $\Sigma$ and $\Lambda$ 
with branching ratios larger than 5 percent.

\begin{figure}[ht] 
\epsfxsize=12.0cm \epsfysize=7.0cm
\centerline{\epsfbox{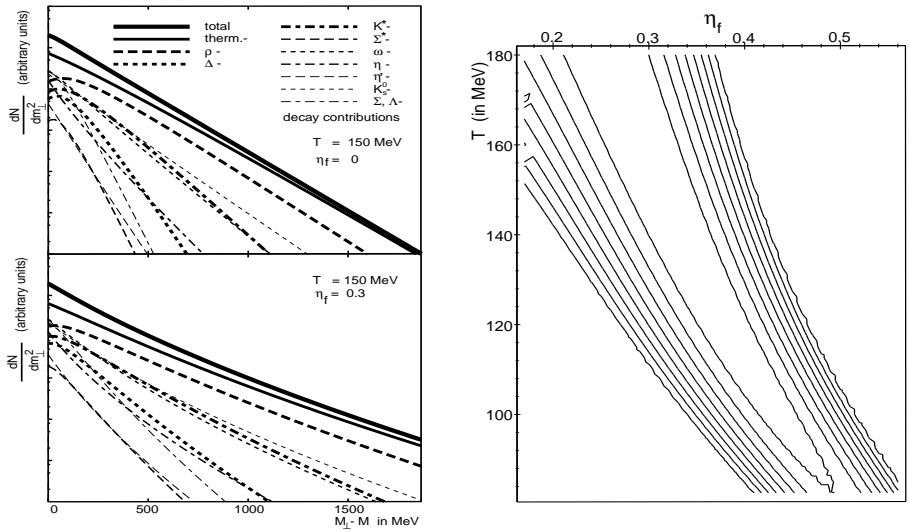} }
\caption{ {\bf Left:} One pion transverse momentum spectra according
to (\ref{7}) for different values of the transverse flow $\eta_f$.
{\bf Right:} The $\sigma$-confidence levels for a fit of our model
to the NA49 $h^-$-spectrum at midrapidity.}
\label{fig1}
\end{figure}

For a given model emission function, both the
two-particle spectrum (\ref{1}) and the one-particle spectrum
$E_p\, dN/ d^3p = \int d^4x\, S(x,p)$ is determined. On the
LHS of Fig.~\ref{fig1}, we plot the rapidity ($y$) integrated
pion transverse mass spectrum
  \begin{equation}
    {dN_\pi\over dm_\perp^2} = \int dy\, \int d^4x\, S(x,p)\, 
    \label{7}
  \end{equation}
for two sets of source parameters without ($\eta_f = 0$) and
with ($\eta_f = 0.3$) transverse flow.
Note that the geometric parameters $R$, $\Delta\eta$, $\tau_0$ of 
the source enter only in the normalization of this spectrum. The 
slope is essentially determined by the temperature $T$ and 
the transverse flow $\eta_f$ and contains no information on the 
source geometry. The direct
pions reflect essentially an effective ``blueshifted'' temperature 
$T_{\rm eff} = T \sqrt{ {1 + \langle \beta_t \rangle \over 
1-\langle\beta_t\rangle}}$. If resonance contributions to (\ref{7}) are
included, these statements change in many details, most notably, 
resonance contributions lead to a low $p_T$-enhancement, and a more
pronounced flattening of the spectra for larger transverse flow.
The main message however remains unaltered:
the unnormalized one-particle spectra (\ref{7}) contain no geometrical
information, their slope determines only a combination of temperature
and transverse flow. This is clearly seen on the RHS of  Fig.~\ref{fig1}
where we show the confidence levels of a fit to the NA49 $h^-$-spectrum
\cite{FOKA}. 

Two-particle correlations are often analyzed in terms of the
Gaussian ansatz
  \begin{equation}
    C({\bf q},{\bf K}) = 1 + \lambda\, {\rm e}^{-R_s^2q_s^2
      -R_o^2q_o^2 -R_l^2q_l^2 -R_{ol}^2q_oq_l}\, .
    \label{8}
  \end{equation}
They allow to disentangle temperature and flow effects. To see this,
one can consider e.g. the approximate expression 
  \begin{equation}
    R_s^2(K_\perp) \approx {R^2\over {1 + (m_\perp/T)\eta_f^2}}\, .
    \label{9}
  \end{equation}
Here, the $m_\perp$-dependence of (\ref{9}) is governed by a prefactor
$\eta_f^2/T$. This indicates that the side radius $R_s(K_\perp)$
may allow to distinguish scenarios with a relative large temperature and a 
small transverse flow size $\eta_f$ from those with a large $\eta_f$ 
and small $T$ which according to the RHS of Fig.~\ref{fig1} 
account for realistic transverse momentum slopes equally well.
We hasten to remark however that (\ref{9}) is obtained in a crude saddle
point approximation neglecting all resonance decay contributions - it is known
to be quantitatively unreliable. This was one of our main motivations
for a quantitatively accurate, numerical calculation~\cite{WH96} of the
two-particle correlation $C({\bf q},{\bf K})$ and its Gaussian widths.

\begin{figure}[ht] 
\epsfxsize=12.0cm \epsfysize=4.0cm
\centerline{\epsfbox{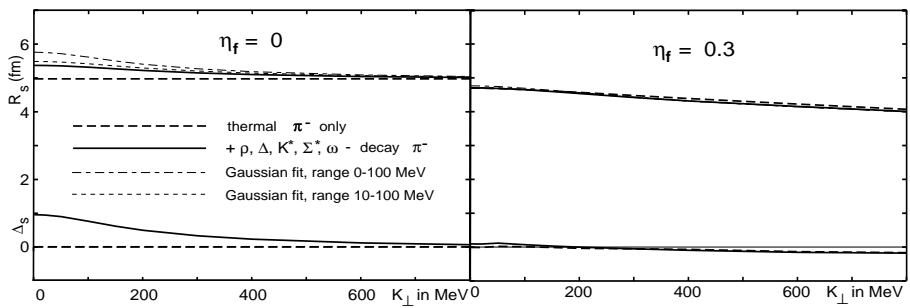} }
\caption{ HBT radius parameters $R_s$ and the kurtosis $\Delta_s$
calculated from the $q$-moments and compared to a Gaussian fit
to (\ref{8}). The two plots depict scenarios without ($\eta_f=0$) 
and with ($\eta_f=0.3$) transverse flow.}
\label{fig2}
\end{figure}

Most notably, resonance decay contributions can lead to deviations of 
the correlator from a Gaussian shape. This is a consequence of the
exponential resonance decay law and can lead to significant quantitative 
ambiguities in analyses based on (\ref{8}), see e.g. the different 
Gaussian fits depicted in Fig.~\ref{fig2}. We hence suggest~\cite{WH96} 
to characterize the correlator by a set of so-called $q$-variances 
which do not depend
on any assumption about the shape. For a unidirectional analysis of the  
correlations $\tilde C(q_i,{\bf K}) \equiv C(q_i, q_{j\ne i}{=}0,{\bf K})$
along the three Cartesian axes, these $q$-moments read
  \begin{eqnarray}
  \label{10}
     R_i^2({\bf K}) &=& {1\over 2\,\langle\!\langle q_i^2 
     \rangle\!\rangle}\, ,
  \\
  \label{11}
     \langle\!\langle q_i^2 \rangle\!\rangle
     &=& {\int dq_i\, q_i^2\, [\tilde C(q_i,{\bf K})-1] \over
          \int dq_i\, [\tilde C(q_i,{\bf K})-1] }\, ,
  \\
  \label{12}
     \lambda_i({\bf K}) &=& (R_i({\bf K})/\sqrt{\pi})
     \int dq_i\, [\tilde C(q_i,{\bf K})-1] \, .
  \end{eqnarray}
Note that for a correlator of Gaussian shape, these expressions coincide 
with those of a fit to (\ref{8}). Still, they are well-defined for 
arbitrary shapes of the correlator. Deviations from a Gaussian shape
can be discussed in terms of higher $q$-moments, most easily in terms
of the ``kurtosis''
  \begin{equation}
  \label{13}
    \Delta_i = { \langle\!\langle q_i^4 \rangle\!\rangle
                 \over
                3\, \langle\!\langle q_i^2 \rangle\!\rangle^2} - 1\, .
  \end{equation}
%
%
Fig.~\ref{fig2} shows the numerically obtained results for the side radius 
if resonance decay contributions are added: $R_s$ develops a 
$K_\perp$-slope even for scenarios without transverse flow. 
The reason is that the resonances propagate outside of the thermally
equilibrated region before decaying. This ``lifetime effect'' 
leads to exponential tails in $S_{R\to\pi}$ and tends to increase the 
Gaussian widths $R_i$. Since the relative abundance of resonances is 
larger for small $K_\perp$, this tail is more pronounced in the region
of small $K_\perp$, thereby leading to a $K_\perp$-slope of $R_s$.
In contrast, for a finite transverse flow $\eta_f = 0.3$ the 
$K_\perp$-dependence of $R_s$ is due to the flow, and resonance decay 
contributions do not change the slope of $R_s$ in our model. This can be
traced back to a shrinking transverse size of the direct 
resonance emission function $S_R^{dir}$ which for the case 
depicted in Fig.~\ref{fig2}, counterbalances the lifetime 
effect almost exactly~\cite{WH96}.

Here, the main message is that resonance decay contributions can lead
to a $K_\perp$-slope of $R_s$, even in scenarios without transverse
flow, i.e., they are faking a signal previously attributed to transverse 
collective dynamics. However, the physical origin of the
$K_\perp$-slope of $R_s$ is different for the two scenarios in
Fig.~\ref{fig2} and it is well understood~\cite{WH96} that this 
physical difference manifests itself in a much more prominent 
``non-gaussicity'' for the case $\eta_f=0$. The kurtosis (\ref{13})
allows for a quantitative measure of deviations from a Gaussian
shape and hence it provides  at least for the class of models discussed here,
the cleanest distinction between scenarios with and without transverse 
flow. We hence expect that in a detailed analysis of two-particle
correlation functions, based on Eqs. (\ref{10})-(\ref{13}), the
effects of temperature and transverse flow can be disentangled most
clearly.

In this short presentation, we have focussed entirely on one aspect,
the distinction between thermal excitation and transverse collective 
dynamics in the collision region. We conclude by pointing out that various
other geometrical and dynamical characteristics of the source are
accessible via the analysis of two-particle correlations. Especially,
we mention information on the longitudinal expansion, emission time
and emission duration as well as the longitudinal and transverse size
of the boson emitting region. In the context of the model (\ref{5}),
resonance decay contributions to these characteristica are quantitatively
under control and we are currently working on the determination of
the model parameters of (\ref{5}) from the data~\cite{FOKA} taken by 
the NA49 Collaboration.

This report is based on joint work with Ulrich Heinz.
It was supported by BMBF, DFG, GSI and DPNC. 


\begin{thebibliography}{999}
\bibitem{S73} 
  E. Shuryak, Phys. Lett. B{\bf 44}, 387 (1973).
\bibitem{CH94} 
  S. Chapman and U. Heinz, Phys. Lett. B{\bf 340}, 250 (1994).
\bibitem{H57} R. Hofstadter, Ann. Rev. Nucl. Sci. {\bf 7}, 231 (1957).
\bibitem{S92} B.R. Schlei e.al., Phys. Lett. B{\bf 293}, 275 (1992);
  ibidem B{\bf 376}, 212 (1996).
\bibitem{HEI96} H. Heiselberg, Phys. Lett. B{\bf 379}, 27 (1996).
\bibitem{WH96}
  U.A.Wiedemann and U. Heinz, nucl-th/9610043 and nucl-th/9611031.
\bibitem{FOKA}
  P. Foka for the NA49 Collaboration, these proceedings.
\end{thebibliography}
\end{document}